\documentstyle[12pt]{article}
\newcommand{\be}{\begin{equation}}
\newcommand{\ee}{\end{equation}}
\newcommand{\bea}{\begin{eqnarray}}
\newcommand{\eea}{\end{eqnarray}}

\title{
Application of symbol-operator technique for effective action
computation }

\author{A.T. Banin$^{1}$\footnote{atb@math.nsc.ru}, N.G. Pletnev$^{1}$\footnote{pletnev@math.nsc.ru}}
\date{{\it ${^{1}}$Institute of Mathematics, \\
Novosibirsk, 630090, Russia,\\
\vspace{0.7cm}}}

\begin{document}

\maketitle
\begin{abstract}
In the present talk we briefly demonstrate an elegant and
effective technique for calculation of the trace expansion in the
derivatives of background fields. One of main advantages of the
technique is manifestly (super)symmetrical and gauge invariant
form of expressions on all stages of calculations. Other advantage
is the universality of the calculation method.
Some particular examples and results are presented.
\end{abstract}
\thispagestyle{empty}

\section{Preface}
Most physical applications are related to computation of a
particular numerical or analytical result for a kernel or a trace of
some operator. One of such problem in QFT is calculation of
one-loop effective action (EA)
\be \Gamma_{1-loop}={i\over 2}{\rm
Tr}\ln (\hat{H}),
\ee
where $\hat{H}$ is the inverse propagator
(in external background fields) for the model under consideration.
In most interesting cases exact calculation of the EA is impossible.
In the same time the
quantum calculations based on operator algebras and therefore
intend calculation of operator products and functions of
operators. Difficulties in such kind of calculation arise because
of noncommutative properties of any operator algebra.
Other problem, which is mostly related to physical models in external
background fields, is the loss of manifest covariance during calculation.

The method presented in this talk pretends to be a universal solution which allows
to lay described above problems aside. It is based on a conception of operator
symbols. Methods of symbol-operator calculus are widely used in the deformation
quantization. Unfortunately computational technique, which are used in the framework
of deformation quantization rather presents purely academical interests then practical.

Shortly speaking, symbol is a function in the phase-space of the considered
model, which can be associated to some operator and verse visa.
Noncommutative star-product operation between symbols is quite analogous to product
of the associated operators.

\section{Basics}
In physics, a proposal to reformulate an analysis in the operator algebras on a
symbol calculus language has been introduced by Berezin \cite{Ber}. A concept of symbol $\sigma(\hat{O})$ of the
operator $\hat{O}$ plays the basic role in this
construction. The symbol is a classical function of finite
number of the phase-space variables $\gamma^{A}$ associated with the operator
$\hat{O}(\gamma)$ which has some defined ordering
(we use the notations $\sigma(\hat{\gamma}_{A})=\gamma_{A}$).

The starting point of the symbol-operator correspondence is an
associative algebra of operators that defines a
noncommutative space and can be described in terms of a set of
operators $\hat{\gamma}_{A}$ and relations
\be\label{alg}
[\hat{\gamma}_{A},\,\hat{\gamma}_{B}]=\hbar\omega_{AB}(\hat{\gamma}).
\ee
In the framework of symbol-operator technique the parameter $\hbar$
plays the role of a non-commutativity parameter.

Let's associate the generators with coordinates $\hat{\gamma}_{A}$
and consider the algebraic structure as the formal power series
in these coordinates modulo to the relations (\ref{alg}).

For example the known {\em Heisenberg-Weyl algebra} with generator set
$\{\hat{\gamma}_{A}\}=\{\hat{P}_{i},\hat{Q}^{j}\}$
is defined by the commutation relation
$[\hat{Q}^{j},\,\hat{P}_{i}]= i\hbar\delta_{i}^{j}$.
An operator family of normal ordered displacement
operators
\be\label{swks}
\hat{\Omega}(p,q)=:{\rm e}^{{i\over \hbar}(q\hat{P}+p\hat{Q})}:,
\ee
forms a complete operator basis for any operator obeying certain
conditions. Colons stand for some normal ordering.
Then we can present some operator $A(\hat{P},\hat{Q})$
in the enveloping Heisenberg algebra in the following way
\be\label{rule}
\hat{A}(\hat{P},\hat{Q}) =
\int {d^n p\, d^n q \over (2\pi)^n} \;
A(p,q)\hat{\Omega}(p,q),
\ee
where coefficient $A(p,q)$ is a smooth function on $T^{*}M$, which is
called the symbol of the operator $\hat{A}(\hat{P},\hat{Q})$.
It should be mentioned that definition of symbol in (\ref{rule})
is straightforwardly depend on the ordering rule in (\ref{swks}).

The illustrated above construction can be extended to a more general case.
In Weyl's quantization procedure (i.e. wholly symmetric ordering) a classical observable $A(\gamma)$, some square integrable
function on phase space $X=(\gamma, \omega)$,  is one-to-one
associated to a bounded operator $\hat{A}$ in the Hilbert space
by the {\em Weyl mapping}
\be\label{wmapin}
\hat{A}= \int_{X}
d\mu(\gamma)\, A_{-s}(\gamma) \hat{\Omega}(\gamma;s).
\ee
Here parameter $s$ encodes all information about the ordering prescription.
An inverse formula which maps an operator into its symbol by {\em
Wigner mapping},  is given by the trace formula
\be\label{wignm}
A(\gamma)= {\rm Tr}(\hat{\Omega}(\gamma;-s)\hat{A}),
\ee
Both  formulae (\ref{wmapin}, \ref{wignm}) are determined by the
choice of the {\em Stratonowich-Weyl kernel}
$\hat{\Omega}(\gamma;s)$, which is the Hermitian operator
family parameterized by the parameter $s$ and constructed from
the operators $\hat{\gamma}_{A}$.
The Stratonowich-Weyl kernel (or a {\em quantizer} and also a
{\em dequantizer}) possesses by a number of properties
(see for example ref. \cite{prop}).
One can see from expressions (\ref{swks}) and (\ref{rule}) that
a symbol can possess a parametric dependence on
$\hbar$ by formal power series.

Correspondence (\ref{wmapin}) relates
$\hat{A}$ to $A_{-s}(\gamma)$ via integration. In practical calculation
it is also helpful to employ a differential form of the relation
$$
\hat{A}=A_{-s}(-i\partial_{\gamma})\hat{\Omega}(\gamma;s)|_{\gamma=0}
$$
We can choose several different rules of normal
ordering for operator products. For instance, the {\em Weyl ordering} (totally symmetrized
operator product) is often a preferred choice for physical
applications since it treats self-adjoint operators
$\hat{P}, \hat{Q}$ symmetrically. This ordering prescription has
specific features leading to a possibility to construct the real symbols
for the operators (i.e. complex conjugation is an algebra
anti-automorphism). Other ordering prescriptions convenient for
practical calculations are the standard $PQ$ (all $\hat{P}$ are
disposed from the left of all $\hat{Q}$) and the antistandard $QP$.

The symbol of the non-commutative product of
operators can be written as a non-local star product
$$
(A\star B)(\gamma)\leftrightarrow\hat{A}\hat{B},
$$
which for a constant Poisson structure is called {\em Moyal product} and
might be treated for a particular case of Weyl ordering
prescription (symmetrical) in integral and differential forms as follows
\bea\label{hstar}
(A\star B)(p,q) &=&
\int {d\xi\,d\eta \over 2\pi\hbar}\,{d\xi'\,d\eta' \over 2\pi\hbar}\;
{\rm e}^{{i \over \hbar}S}
A(\xi,\eta)B(\xi',\eta'),\nonumber \\
(A\star B)(p,q)&= &A(p, q)
{\rm e}^{i\hbar(
{\stackrel{\leftarrow}{\partial} \over \partial q}
{\stackrel{\rightarrow}{\partial} \over \partial p}-
{\stackrel{\leftarrow}{\partial} \over \partial p}
{\stackrel{\rightarrow}{\partial} \over \partial q})
}
B(p,q),
\eea
where
$
S=\det \pmatrix{1 & 1 & 1\cr q& \xi& \xi'\cr p&\eta&\eta'\cr}.
$

Of course, the non-locality of a star product is a consequence of the
ordinary quantum-mechanical non-locality.

In the phase space parameterized by coordinates
$\gamma$ we have an analogue of algebra (\ref{alg}) with
the {\em Moyal bracket}
\be\label{algmoy}
[A,\,B]_{\rm MB} =
A\star B-B\star A=
i\hbar \{A,B\}+O(\hbar^2),
\ee

So far the star product is defined only by  the
relations (\ref{alg}). Unfortunately, all these
results are formal in the sense that they do not offer a
receipt for the procedure of the star-product construction.
The basic problem in attempt to generalize the exponentiation idea
(\ref{hstar})
to a non-constant Poisson structure is that
$\stackrel{\leftarrow}{\partial},\,\stackrel{\rightarrow}{\partial}$
no longer commute with the $\omega$. Nevertheless, recall that for
$\hat{A},\hat{B}$ in some Lie algebra and for
$\exp (\hat{A}) \exp (\hat{B})= \exp (\hat{C})$ the
Campbell-Baker-Hausdorff formula allows to define
$\hat{C}$ as a formal series whose terms are elements in
the Lie algebra generated by $\hat{A},\hat{B}$.
The associativity of such a way defined star product is induced
from the associativity of the group multiplication.

\section{Construction}
Definitions and discussion presented in the previous section  might have
remained just a toy for mathematicians if they didn't have a powerful
practical application.
For our purposes, the most important formula, which takes place
in the framework of the symbol calculus due to the special
properties of the Stratonowich-Weyl kernel, is the trace definition
\be\label{symtr}
{\rm Tr}\hat{A} = \int_{X}d\mu(\gamma)\,A(\gamma),
\ee
where $X=(\gamma, \omega)$ is the phase space with invariant measure
$d\mu$ and $A(\gamma)$ is some symbol of operator $\hat{A}$.
It shold be noted that the definition (\ref{symtr})
is correct for any allowed choice of the Stratonowich-Weyl kernel
and for any ordering prescription in the enveloping algebra (\ref{alg}).
According to Eq (\ref{symtr}) finding of trace of the
operator $\hat{A}$ is reduced to constructing the corresponding
symbol $A(\gamma)$.
For example a symbol of a some operator function would be written as a Taylor
series
$$
\sigma(f (\hat{A}))=\sum_{n} {1 \over n!} f^{(n)}\,A\star A\star\ldots\star A,
$$
and to find the trace of the corresponding operator one have
to perform the integration of this symbol according to (\ref{symtr}).  However, for the
operators having intricate structure, obtaining its symbol via
explicit evaluation of Wigner's mapping appear intractable.
Moreover, the star product formulae (\ref{hstar}) are
very unhandy for practical computations. But, since the differential
form involves an exponential of derivative operators, it may be evaluated
through translation of the function arguments. This can be easily seen
from (\ref{hstar}), which might be rewritten in the following form
\be
(A\star B)(p,q)= A(p-{i\over 2}\hbar{\partial \over \partial q_1},
q+{i\over 2}\hbar{\partial \over \partial p_1})
B(p_1,q_1)|_{p_1=p, q_1=q}=
A(p_{\hbar},q_{\hbar})B(p_{\hbar},q_{\hbar})\times 1,
\ee
where quantities
\be\label{hrep}
p_{\hbar}=p-{i\over 2}\hbar\stackrel{\rightarrow}{\partial_q} \,
q_{\hbar}=q+{i\over 2}\hbar\stackrel{\rightarrow}{\partial_p}
\ee
were introduced. They should be considered as a special right regular
representation of generating operators $\hat{P}, \hat{Q}$.

In the general case, we can rewrite expression (\ref{hstar}) in a
more symmetrical form
\be\label{rep}
(A\star B)(\gamma) =
 A(\stackrel{\rightarrow}{\gamma_{\hbar}})
 B(\stackrel{\rightarrow}{\gamma_{\hbar}})\times 1 =
  1\times A(\stackrel{\leftarrow}{\gamma_{\hbar}})
 B(\stackrel{\leftarrow}{\gamma_{\hbar}}).
\ee

Operators
$
\stackrel{\leftarrow}{\gamma}_{\hbar}, \,
\stackrel{\rightarrow}{\gamma}_{\hbar}
$ are left and right regular representation operators
$\hat{\gamma}$ with commutation relations (\ref{alg}).

As it can be seen from (\ref{hrep}) that the operators $A_{\hbar}$ can
be obtained from adjoint action of some shifting operator $A_{\hbar}=U^{-1}AU$.
It should be noted that the recipe
\be\label{trans}
A \rightarrow A_{\hbar}=U^{-1}AU
\ee
in the geometric quantization aspect is just a {\em prequantization procedure}
for the phase-space $X=(\gamma,\omega)$. Really, prequantization on $X$ proceed
by finding a linear map from Poisson algebra of smooth function $f$ to an algebra
of operators $\hat{f}$. The recipe is $\hat{f}=f-i\hbar\nabla_{X_{f}}$, where $X_{f}$ is
Hamiltonian vector field of $f$ (see for example Ref. \cite{jose})

Particulary, for the representation (\ref{hrep}) operator $U$ has simple from
\be\label{exu1}
U={\rm e}^{{\hbar\over 2}(\stackrel{\leftarrow}{\partial}_{q}\stackrel{\rightarrow}{\partial}_{p}-\stackrel{\leftarrow}{\partial}_{p}\stackrel{\rightarrow}{\partial}_{q})}.
\ee

The special representation of operators has very important properties:
the relation between symbols and operators which is inverse to (\ref{trans})
can be written as follows:
\be\label{bridge}
A = \stackrel{\rightarrow}{A_{\hbar}}\times 1 = 1\times\stackrel{\leftarrow}{A_{\hbar}}
\ee
This rule just means that all derivatives that act on nothing
must be omitted, i.e. for (\ref{hrep}) we get $(p_{\hbar},\,q_{\hbar})\times 1 = (p,\,q)$

Expressions (\ref{trans}) and (\ref{bridge}) are mutually inverse. If we note
$Q(f)=UfU^{-1}$ and $Q^{-1}(f)=1\times f\times 1$, then star-product (\ref{rep})
between two function on the phase space can be written in the following form
\be
A\star B =Q^{-1}(Q(A)\circ Q(B)),
\ee
where $\circ$ is a noncommutative operator product, which is undoubtedly associative.
It is not difficult to see that if the presented above construction has direct relation
to the well-known Fedosov construction (see Ref. \cite{Fed}). The representation
$A_{\hbar}$ is the flat section of Fedosov connection corresponding to function $A$.

As we can see, the described construction is uniquely determined by the choice of the
operator $U$. The form of operator $U$ for different cases is widely discussed
by many authors (see for example Ref. \cite{kar}).
In all known cases the operator $U$ has the simple exponent form and can expressed
via functions $\omega_{AB}(\gamma)$ from relations (\ref{alg})
The explicit form of operator $U$ is determined by the chosen ordering
prescription. In the applications, the form of the operator $U$ is often chosen for reasons
of practical needs.

According to (\ref{rep}), there are equivalent left and right special representations.
The application of transformation rule (\ref{trans}) depends on chosen representation. If one
decided to use right regular representation the right derivatives in operator (\ref{exu1}) must
be considered as "passive", i.e. as parameters (or as coordinates in the tangent space).

It was tested by practical calculation (see ref. \cite{we}) that this construction can be easily
generalized to any form operator algebra (\ref{alg}), including
algebra od supersymmetry.

How we can use this representation in practice? For example we need to calculate
a trace of some operator function $f(A(\hat{\gamma}))$. The recipe is given:\\
1. Choose transformation operator $U$;\\
2. Find special representation $\gamma_{\hbar}$ of all operators from the basis $\hat\gamma$;\\
3. Replace all operators in $f(A(\hat{\gamma}))$ by their representation, i.e. $f(A(\gamma_{\hbar}))$;\\
4. Drop all derivatives after moving them to the one side;\\
5. Integrate the expression (symbol) over the phase-space.\\

The main advantages of the method can be seen from the recipe. Firstly, we have wholly
covariant from of the expression. Secondly, we can expand the expression
up to the interesting us degrees of a noncommutativity parameter,
using the noted earlier methods.

\section{Examples}
To clarify the structure of the construction and to show the simplicity we
consider several examples.

\subsection{Curved phase space}

For the electrodynamics we have the following commutation relations:
\be
[\nabla_{\nu}, x^{\mu}]= \delta_{\nu}^{\mu}, \quad [\nabla_{\mu},\nabla_{\nu}]=F_{\mu\nu}(x).
\ee
Let $ip_{\mu}=\sigma ({\partial\over \partial x^{\mu}}), x^{\nu}=\sigma (x^{\nu})$ are notation for the symbols.
Then symbol for covariant derivatives will be $(ip + A(x))_{\nu}=\sigma (\nabla_{\nu})$
We have to choose the ordering prescription for the covariant derivatives. For
$pq$-ordering prescription and symmetrical ordering prescription for covariant derivatives we can
write the following transformation operator
$
U={\rm e}^{-i{\stackrel{\leftarrow}{\partial}\over \partial p_{\mu}}\stackrel{\rightarrow}{\nabla_{\mu}}}.
$
The operators in right regular representation have the forms
\bea
x^{\mu}_{\hbar}&=&x^{\mu}+i{\stackrel{\rightarrow}{\partial}\over \partial p_{\mu}},\nonumber\\
\nabla^{\hbar}_\mu &=&U^{-1}
( ip_\mu + A_\mu)U =
ip_{\mu}+ i\int^1_0 d \tau\, \tau F_{\nu\mu} (x^{\rho} + i\tau{\partial\over \partial p_\rho}){\partial\over \partial p_\nu} =\nonumber\\
&=&ip_\mu + {i\over 2}  F_{\nu\mu}{\partial\over \partial p_\nu} -
{1\over3}
F_{\nu\mu,\lambda}{\partial^{2}\over \partial p_\lambda \partial p_\nu} - {i\over 8}
F_{\nu\mu,\lambda\sigma}{\partial^{3} \over \partial p_{\sigma}\partial p_{\lambda}\partial p_{\nu}} + \ldots
\eea
We have obtained nothing but a representation of the vector potential in the Fock-Schwinger gauge
$$
A_\mu(q)=q^\nu\sum^\infty_{n=0}{1\over n + 2}{1\over n!}
x^{\alpha_1}... x^{\alpha_n} F_{\nu\mu,\alpha_1 ...\alpha_n},
$$
 without explicit solving the gauge condition $x^\mu A_\mu=0.$
The symbols of the operators can be found according to (\ref{rule}).
The measure in the phase-space for (\ref{symtr}) is $d\mu=d^{4}x~d^{4}p$.

\subsection{Flat {\cal N}=1 superspace}

For the supersymmetry models the starting point is the algebra of supersymmetry
\be\label{salg}
[\partial_{\nu}, x^{\mu}]= \hbar\delta_{\nu}^{\mu}, \quad \{D_{\alpha},\bar{D}_{\dot{\beta}}\}=i\partial_{\alpha\dot{\beta}}\quad
\{D_{\alpha}, \theta^{\beta}\}=\delta_{\alpha}^{\beta},\quad \{\bar{D}_{\dot{\alpha}}, \bar{\theta}^{\dot{\beta}}\}=\delta_{\dot{\alpha}}^{\dot{\beta}}.
\ee
We will write symbols for differential operators as follows
\be
\sigma (D_{\alpha})=\psi_{\alpha},\quad \sigma (\bar{D}_{\dot{\beta}})=\bar{\psi}_{\dot{\beta}},\quad\sigma ({\partial \over \partial x^{\alpha\dot{\beta}}})=p^{\alpha\dot{\beta}}.
\ee
The transformation operator can be chosen in the form
\be
U={\rm e}^{1/2\theta p\partial_{\bar\psi}-1/2 \partial_\psi
p \bar\theta}{\rm e}^{\partial_\psi D_{\theta}+
\partial_{\bar\psi}\bar{D}_{\bar{\theta}}}{\rm e}^{i\partial_p \cdot \partial_x}.
\ee
Such a form of operator leads to the following form of operator representation:
\bea
&\psi_{\hbar}=\psi-{1\over 2}\partial_{\bar\psi} p,\quad
\bar{\psi}_{\hbar}=\bar\psi-{1\over 2}\partial_\psi p,&\nonumber\\
&x_{\hbar}= x +i\partial_p,\quad
\theta_{\hbar}=\theta + \partial_\psi,\quad
\bar{\theta}_{\hbar}=\bar\theta +\partial_{\bar\psi}&\nonumber
\eea
where we use short forms of derivatives $\partial_\psi={\partial\over \partial\psi}, \partial_{\bar\psi}={\partial\over \partial{\bar\psi}}$
etc. It is easy to see that operators in such representation satisfy algebra (\ref{salg}).

The superspace measure for (\ref{symtr}) is $d\mu=d^{4}x~d^{4}p~d\theta~d\bar{\theta}~d\psi~d\bar{\psi}$.

\subsection{Curved {\cal N}=1 superspace}

The algebra of covariant derivatives has a form
\be
\{\nabla^{\hbar}_\alpha,\nabla^{\hbar}_{\dot\alpha}\}=
i\nabla^{\hbar}_{\alpha\dot\alpha},
\quad
[\nabla^{\hbar}_{\alpha\dot\alpha},\nabla^{\hbar}_{\beta\dot\beta}]=
i(C_{\dot\beta\dot
\alpha}f_{\beta\alpha}+C_{\beta\alpha}f_{\dot\beta\dot\alpha}),
\ee

$$
[\nabla^{\hbar}_{\dot\beta},\nabla^{\hbar}_{\alpha\dot\alpha}]=
C_{\dot\beta\dot\alpha}W^{\hbar}_\alpha, \quad
[\nabla^{\hbar}_{\dot{\alpha}},W^{\hbar}_{\alpha}]=0,
$$
where
$W^{\hbar}_\alpha =W_\alpha +  \partial^{\beta}
f_{\beta\alpha}-i\partial_\alpha D'$.
The quantities $f$, $D'$ are the standard notation for superfields
$f_{\alpha\beta} = {1\over 2} \nabla_{(\alpha} W_{\beta)},$
$D'=- {i \over 2} \nabla^\alpha W_\alpha,$
$\nabla^\alpha W_\alpha + \bar\nabla^{\dot\alpha} \bar W_{\dot\alpha} =0$.
The transformation operator can be chosen as follows:
\be
U={\rm e}^{-\bar{\partial}_{\bar{\psi}}\bar{\nabla}} {\rm e}^{{1 \over
2}(\partial_{\psi} p \bar{\theta}-\theta p
\bar{\partial}_{\bar{\psi}})} {\rm e}^{-\partial_{\psi}\nabla}
{\rm e}^{-i\nabla\partial_{p}}.
\ee
For the presented example, the expansion of the derivatives operators
are
\be
\nabla^\hbar_\alpha=
\psi_\alpha-{1\over 2}\bar\partial^{\dot\alpha}p_{\alpha\dot\alpha}+
{i\over 4}\bar\partial^{\dot\alpha}
(\partial^{\dot\beta}_\alpha f_{\dot\beta\dot\alpha}+
\partial^{\beta}_{\dot\alpha} f_{\beta\alpha})-
{1\over 3}\partial_\alpha\bar\partial^{\dot\alpha}i\bar W_{\dot\alpha},
\ee
$$
+{1\over 3}\bar\partial^2 iW_\alpha+{1\over 4}\partial_\alpha\bar\partial^2
D'+{3\over 4!}\bar\partial^2\partial^{\beta}if_{\beta\alpha}+...
$$

$$
\nabla^\hbar_{\dot\alpha}=
\psi_{\dot\alpha}-{1\over 2}\partial^\alpha p_{\alpha\dot\alpha}+
{i\over 4}\partial^\alpha
(\partial^{\dot\beta}_\alpha f_{\dot\beta\dot\alpha}+
\partial^{\beta}_{\dot\alpha} f_{\beta\alpha})
+{1\over 3}\partial^2 i\bar W_{\dot\alpha}
$$
$$
-{1\over 3}\bar\partial_{\dot\alpha}\partial^\alpha iW_\alpha
-{1\over 4}\bar\partial_{\dot\alpha}\partial^2 D'
+{3\over4!}\partial^2\bar\partial^{\dot\beta}if_{\dot\beta\dot\alpha}
+...,
$$
Similarly, for a vector derivative we have
\be
\nabla^{\hbar}_{\alpha\dot\alpha}=
i p_{\alpha\dot\alpha}+{1\over 2}(\partial^\beta_{\dot\alpha}f_{\alpha\beta}
+\partial^{\dot\beta}_{\alpha} f_{\dot\alpha\dot\beta})+
\partial_\alpha\bar W_{\dot\alpha}
+\bar\partial_{\dot\alpha} W_\alpha+
{1\over 2}(\partial_\alpha\bar\partial^{\dot\beta}
f_{\dot\beta\dot\alpha}
+\bar\partial_{\dot\alpha}\partial^{\beta}f_{\beta\alpha})+i
\partial_\alpha\bar\partial_{\dot\alpha}D'.
\ee
All derivatives in above expressions are right and act through.
Here the notations mean $\partial^{\alpha}={\partial\over\partial \psi_{\alpha}}, \partial_{\dot\alpha}={\partial\over\partial \bar\psi^{\dot\alpha}}$ and
$\psi^{\alpha},\bar\psi_{\dot\alpha}$ are the grassmanian coordinates in the momentum superspace.
The super phase-space measure is coincide with given for the flat superspace.

\section{Results}

In this section we present the results which has been obtained
by the method. This method was applied for calculating traces of evolution operators
and one-loop effective action (see Refs. \cite{we}).

It is well-known that the evaluation of the determinants of the (pseudo)differential
operators always involves some kind of regularization. For actual computations
of the effective action we use elegant $\zeta$-function regularization formulated directly in
(super)space. Within this regularization scheme the functional
determinant
$ {\rm Det}_{\zeta}(\hat{H})=\exp
(-\zeta'_{\hat{H}}(0)) $
is preserve the symmetry and the gauge invariance. As
a result the effective action looks like
$ \Gamma_{(1)}=-{i \over 2}\zeta '_{\hat{H}} (0) $
and $\zeta$-function is defined by the
expression
\be\label{zeta}
\zeta_{\hat{H}} (s)= {1 \over \Gamma
(s)}\int_{0}^{\infty}dT\,T^{s-1}{\rm Tr} ({\rm e}^{-{T\over
\mu^2}\hat{H}}) = {1 \over \Gamma (s)}\int_{0}^{\infty}dT\,T^{s-1}
\int dz \,K({T\over \mu^2}),
\ee
where $dz$ is an appropriate
(super)space measure and $\mu$ is a renormalization point, which
is introduced to make $T$ dimensionless. The
quantity $K(T)$ is the coincidence limit of the heat kernel which can
be represented in form of Schwinger-DeWitt expansion over the proper
time parameter $T$
\be\label{deck} K(T)= {1 \over (4\pi
T)^2}\sum_{f=0}^{\infty}a_{f}T^{f}.
\ee
Here $a_{f}$ are DeWitt-Seely coefficients which are the scalars constructed from
the coefficients of the operator $\hat{H}$.

\subsection{Massive scalar field}

As the first example we shall consider a massive scalar field theory with
the lagrangian
$$
{\cal L} = {1 \over 2}\partial_{\mu} \phi \partial^{\mu} \phi -
m^2\phi^2 - U(\phi).
$$
The heat kernel for this model after replacement of the operators
has a form
$$
K(T)=
\int {d^4 p\over (2\pi)^4} {\rm e}^{-T (p^2_{\hbar} + V (x_{\hbar}))}.
$$
Then two separate problems can be studied: the problem of the decomposition
over derivatives degrees and the problem of fixed degree of the
potential $V=m^2+U''$.

If we leave in decomposition only fourth order terms in
derivatives, after rearranging by extracting full derivatives, we
obtain the known result \cite{Fli}
\bea
K(T)={1\over (4\pi T)^2}
{\rm e}^{-VT}\times \phantom{vvvvvvvvv}\\ \nonumber \times\left(1-
{1\over 12} T^3 V_\mu V_\mu + {1\over 5!} T^4 V_{\mu\nu}
V_{\mu\nu} - {T^5\over 3\cdot 4!} V_{\mu\nu} V_\mu V_\nu +
{T^6\over 12\cdot 4!} V_\mu^2 V_\nu^2 \right).
\eea
As the another problem we consider
the second order $V$ term. After simple manipulation we obtain
\be
K(T) = \int_{0}^{T}ds_{1}\int_{0}^{s_{1}}ds_{2}V(x)
e^{(s_{1}-s_{2})(\Box+2ip\nabla)}V(x)
e^{-(s_{1}-s_{2})(\Box+2ip\nabla)}.
\ee
This leads to the expression
\be
\Gamma_{(1)} \sim -{1 \over 2}\int {dT \over T} e^{-m^2T}
\int d^4 x {1 \over (4\pi T)^2}{T^2 \over 2 }
V\gamma(T \Box)V,
\ee
where the formfactor has the known representation
\be
\gamma(T \Box) = \int_{0}^{1}ds e^{{1- s^2 \over 4}T \Box}
\ee

\subsection{Non-abelian gauge fields}

We consider of a scalar loop in the external nonabelian field
with the lagrangian
$$
{\cal L}={1 \over 2}\nabla_{\mu} \phi \nabla^{\mu} \phi + m^2 \phi^2;
\quad \hat{H}=-\nabla^{\mu}\nabla_{\mu} + m^2.
$$
According to the prescription described above, we use a $\zeta$-function representation
for one-loop EA with
$H=\nabla^{\mu}_{\hbar}\nabla_{\hbar}^{\mu}+m^2$, where
$\nabla^{\mu}_{\hbar}$ is a covariant pseudodifferential operator.

Further, performing trivial calculations
 and using the Bianchi identities we will
get the known result \cite{Nov} for on-shell background fields, i.e. $F_{\mu\nu,\,\nu}=0$
\bea
K(T)&=&{ {\rm e}^{-Tm^2} \over (4\pi T)^2} (1 + {T^2\over 12} F_{\mu\nu}
F_{\mu\nu} + T^3({F^3\over 180} )+\nonumber\\
&+&{ T^4 \over 2\cdot 144} \left(( F_{\nu\mu} F_{\nu\mu})^2 + {1 \over 5}
 \{F_{\alpha\mu}, F_{\beta\mu} \}^2 +
 {1 \over 7} [F_{\alpha\beta}, F_{\mu\beta}]^2 +
 {1 \over 70} [F_{\alpha\mu}, F_{\rho\sigma}]^2
 \right)),
\eea
where $F^3=F_{\mu\nu} F_{\nu\alpha} F_{\alpha\mu}$. Thus the first
(after  unit) term of the decomposition is related to renormalization
of charge. It should be noted that the huge number of
 terms in the decomposition can be omitted at once,
that essentially reduces work and demonstrates that
computation of higher power corrections might be considerably simplified.

\subsection{Wess-Zumino model}

In this section we demonstrate how to apply the proposed technique to
calculation one-loop EA for the supersymmetrical theories in
the superfield approach.
The doubtless advantage of the offered method is that
this method does not require determinations of many various
Green functions for calculation of the functional trace of the
appropriate heat kernel.  To show it, we obtain the known
K\"ahlerian potential of Wess-Zumino model
and lowest order nonk\"ahlerian contributions to one-loop
effective potential.

The Wess-Zumino theory described by the action
$$
S(\phi,\bar\phi)=\int d^8 z\bar\phi\phi+\int d^6 z
\left({m\phi^2\over 2}+
{g\over 3!}\phi^3 \right) + h.c.
$$
is a good model for test of various supersymmetric methods,
since it has all specific peculiarities of the theories with
chiral fields.

It is convenient to introduce unlimited superfields
$\phi=\bar D^2\psi$ and $\bar\phi=D^2\bar\psi$. In principle this
introduce a new gauge invariance into the action, but in the absence
of background gauge fields, the ghost associated with this gauge
fixing are decoupled.  The functional integration over $\psi, \bar{\psi}$ leads to a
determination of the effective action in the form
$
-{i\over 2}Tr\ln(\hat{H}(x,\theta,D))
$
with the kinetic operator for the given model
\be\label{ham}
\hat{H}=\pmatrix{\lambda&\bar D^2\cr D^2&\bar\lambda\cr} \pmatrix{\bar
D^2&0\cr 0& D^2\cr},\quad\mbox{ where }\quad
\lambda=m+g\phi_{(BG)}.
\ee

Limiting ourselves to a problem of calculation of the first
correction to the potential in decomposition over Grassmanian derivatives, we
can find the K\"ahlerian potential
\be
K^{(1)}=\int{d^4 p\over (2\pi)^4}{1\over 2p^2}\ln(1+{\lambda\bar\lambda\over
p^2}).
\ee
Also the known result for the non K\"ahlerian
terms (auxiliary fields potential) \cite{Pick}, leading to quantum deformations of
classical vacuum of the theory
\be
F^{(1)}={1\over 3\cdot
2^7}\,{D^{\alpha}\lambda D_{\alpha}\lambda \bar{ D}^{\dot\alpha}
\bar{\lambda} \bar{D}_{\dot\alpha}\bar{\lambda} \over
\lambda^2\bar{\lambda}^2}
\ee
can be found.
Certainly, such quantum corrections are important in ${\cal N}=1,2$
supersymmetrical models, since they lead potentially  to the
removal of degeneration in classical vacua of the theory.

\subsection{Heisenberg-Euler lagrangian in SQED}

Now we consider develop manifestly supersymmetrical gauge invariant
strategy of calcu\-lations of one-loop effective action for the most
general renormalizable $N=1$ models including Yang-Mills fields and
chiral supermuliplets
$$
{\cal S}={\rm tr} \int d^6 z\, W^2 +
\int d^8 z\, \bar{\Phi}e^{V}\Phi + [\int d^6\, z P(\Phi)+ h.c.]
$$
In more detail we consider the one-loop diagrams only with external abelian
superfields and the expansion in terms of spinor covariant
derivatives of superfields $W, \bar{W} $ which can not be reduced to
usual space-time derivatives. This approximation corresponds to
generalization of the Heisenberg-Euler Lagrangian of usual QED.

The results coincides with the result of
\cite{Shiz}, obtained by essentially different methods.
\be\label{sgam}
\Gamma_{(1)}= \int d^8z W^2 \int_{0}^{\infty}{dT \over T}{\rm e}^{-Tm^2}
{\cosh(TD') - \cosh(T {\cal H}_{-}) \over {D'}^2-{\cal H}_{-}^2}K(T)_{Schw},
\ee
where ${\cal H}_{-}$ is defined as follows
\be
{\cal H}_{\pm}= (\lambda_{1}\pm i\lambda_{2})^2 =
{1 \over 2}(f^2 \mp i f^*f),
\ee
and
\be
K(T)_{Schw}= {1\over (4\pi T)^2} [\det {fT\over \sinh (fT)}]^{1/2}.
\ee

In the last example we will consider contributions from only the
quantum gauge field $V$. After splitting the field into a
background and quantum part, the quadratic over vector field SYM
action in Fermi - Feynman gauge is
$$
S = -{1 \over 2g^2}{\rm Tr}\left(V[\Box
-iW^{\alpha}\nabla_{\alpha} -
i\bar{W}^{\dot{\alpha}}\bar{\nabla}_{\dot{\alpha}} ]V\right).
$$
Performing the calculation we can find the final result
\be K(T)=
W^2\bar{W}^2\det({{{\rm e}^{TN}-1}\over N}) \det({{{\rm
e}^{T\bar{N}}-1}\over \bar{N}}){1\over(4\pi T)^2}[\det{TF\over
\sinh(TF)}]^{1/2},
\ee
where $N^\beta_\alpha=iD_\alpha W^\beta,$
$\bar
 N^{\dot\beta}_{\dot\alpha}=i\bar D_{\dot\alpha}\bar{W}^{\dot\beta}$.
Such Heisenberg-Euler lagrangians in the superfield form,
which involve higher derivatives, are important for analysis of quantum
structure  ${\cal N}=2,4$ superconformal actions \cite{Buch}.

\section{Resume}
In the present talk it was shortly demonstrated key ideas of the elegant and effective
technique for calculation of the trace expansion in the derivatives of
background fields.

Some particular examples and result were presented.
They demonstrate universality of the proposed technique.

\section{Acknowledgments}
Authors are extremely grateful to Dr. M.Bordag for his invitation to the
Workshop and for the financial support, which made possible this participation.
Author A.Banin is also grateful to Dr. A.Ernst for his assistance.
The work was supported by INTAS grant, INTAS-00-00254 and
by RFBR 99-02-17211, RFBR 00-15-96691 grants.

\end{document}